\documentclass[aps,pre,twocolumn,showpacs,floatfix]{revtex4}

\usepackage{dcolumn}
\usepackage{amssymb}
\usepackage{graphicx}

\begin{document}

\newcommand \be {\begin{equation}}
\newcommand \ee {\end{equation}}
\newcommand \bea {\begin{eqnarray}}
\newcommand \eea {\end{eqnarray}}
\newcommand \nn {\nonumber}
\newcommand \la {\langle}
\newcommand \rl {\rangle_L}

\title{On-off intermittency over an extended range of control parameter}
\author{Eric Bertin}
\affiliation{Universit\'e de Lyon, Laboratoire de Physique,
ENS de Lyon, CNRS, 46 all\'ee d'Italie, F-69007 Lyon, France}

\date{\today}

\begin{abstract}
We propose a simple phenomenological model exhibiting on-off intermittency
over an extended range of control parameter.
We find that the distribution of the 'off' periods has as a power-law tail
with an exponent varying continuously between $-1$ and $-2$, at odds with standard
on-off intermittency which occurs at a specific value of the control parameter,
and leads to the exponent $-3/2$.
This non-trivial behavior results from the competition between
a strong slowing down of the dynamics at small values of the observable,
and a systematic drift toward large values.
\end{abstract}


\maketitle

Numerous non-equilibrium systems exhibit an irregular dynamics, in which quiet periods
alternate with bursts of activity, giving rise to on-off intermittency.
This phenomenon has been characterized experimentally
in many different systems, ranging from electronic devices \cite{Hammer94},
spin-wave instabilities \cite{Rodelsperger95},
plasmas \cite{Feng98}, liquid crystals \cite{John99,Vella03},
multistable fiber laser \cite{Huerta08},
and nanoscopic systems such as nanocrystal quantum-dots \cite{Brokmann03},
or diffusing nanoparticles \cite{Zumofen04}.
A standard characterization of the intermittency phenomenon is the distribution of time spent in the off-state.
In most cases, this distribution exhibits a power-law regime with an exponent close to $-1.5$, with a cut-off at large times
\cite{Hammer94,Rodelsperger95,Feng98,John99,Vella03,Huerta08,Brokmann03,Zumofen04}.
From a theoretical viewpoint, on-off intermittency is
understood as resulting from the effect of multiplicative noise on a system close to
an instability threshold, generating an effective instability threshold that
fluctuates over time, and leading to the alternation of stable (off, or inactive)
and unstable (on, or active) periods.
According to this scenario \cite{Fujisaka1985,Platt1993,Heagy1994,Aumaitre},
the distribution of the duration of the off-periods
has a power-law tail with a well-defined exponent $-3/2$. This
exponent value has a simple interpretation in terms of first-return time statistics:
on a logarithm scale, the dynamics can be mapped onto a random walk
on a half axis (the one corresponding to small values of the physical variable),
so that the duration of off-periods corresponds to the first return time
to the origin of a random walk, whose statistics is known to
behave with a tail exponent $-3/2$ \cite{Feller}.

In spite of the large body of results corroborating this scenario,
some experimental systems have been found to exhibit a different behavior,
with exponent values between $-1$ and $-2$ \cite{Kuno01,Kuno03,Divoux}.
In addition, it has also been observed experimentally in some cases that the 
power-law distributions of the off-periods appear for
an extended range of control parameter \cite{Divoux}, and not only close to a
given threshold.
While a possible explanation of the variations of the exponent might be the presence
of correlations in the dynamics \cite{fractal-noise95,fractal-noise96},
no theoretical scenario is, to our knowledge, presently able to account for the existence
of on-off intermittency over a finite range of control parameter.

In this note, we study both analytically and numerically a simple
stochastic model exhibiting on-off intermittency, with a continuously
varying tail exponent, over an extended range of control parameter.
We also provide a simple interpretation for the upper and lower bounds of this range,
and give an analytical argument, based on random walk properties, to determine the tail
exponent as a function of the control parameter.
Intermittency results in this case from a competition
between a slowing down at small values of the observable, and an
average drift toward larger values.

{\it Model and dynamics.}--
We consider a simple model describing the stochastic dynamics of a continuous positive
variable $x$, corresponding to an arbitrary physical observable. The dynamics is defined
as follows. Starting from a value $x$, the process can jump to a value
in the interval $[x',x'+dx']$ with a probability per unit time
$W(x'|x)\, dx'$, where $W(x'|x)$ is given by
\be \label{def-rates}
W(x'|x) = \nu_0 \rho(x')\, \psi\left(\frac{x'}{x}\right),
\ee
$\nu_0$ being a characteristic frequency.
The distribution $\rho(x')$, normalized according to
$\int_0^{\infty} \rho(x')\, dx' = 1$,
characterizes the a priori probability to choose the value $x'$
(it may thus be thought of as a density of states),
while $\psi(u)$ describes an acceptance rate of the transition
from $x$ to $x'$.
This rate is chosen to be a function of the ratio $x'/x$, so that the process
can be considered as multiplicative.
The density $\rho(x)$ is assumed to behave for small values of $x$
as a power law, namely
\be \label{rho-asympt}
\rho(x) \sim c\, x^{\gamma_0-1}, \qquad x\to 0^+,
\ee
where $\gamma_0$ and $c$ are positive constants. 
For reasons that will appear clear later on, the function $\psi(u)$ is
assumed to fulfill the symmetry
\be \label{def-psi}
\psi\left(\frac{1}{u}\right) = u^{\gamma}\, \psi(u), \quad u>0,
\ee
with $\gamma>0$ a parameter of the model.
When an explicit form of $\psi(u)$ is required, we use in the following
\be \label{def-psi-ex}
\psi_{\mathrm{ex}}(u) = \left\{\begin{array}{c}
u^{\eta}, \qquad \qquad \;\, 0<u<1 \\
u^{-(\gamma+\eta)}, \qquad \qquad u>1
\end{array}\right. 
\ee
($\eta \ge 0$), which fulfills the symmetry property Eq.~(\ref{def-psi}).
In what follows, we consider $\gamma_0$
and $\eta$ as fixed parameters,
and $\gamma$ as the control parameter of the system.

\begin{figure}[t]
\centering\includegraphics[width=7.5cm,clip]{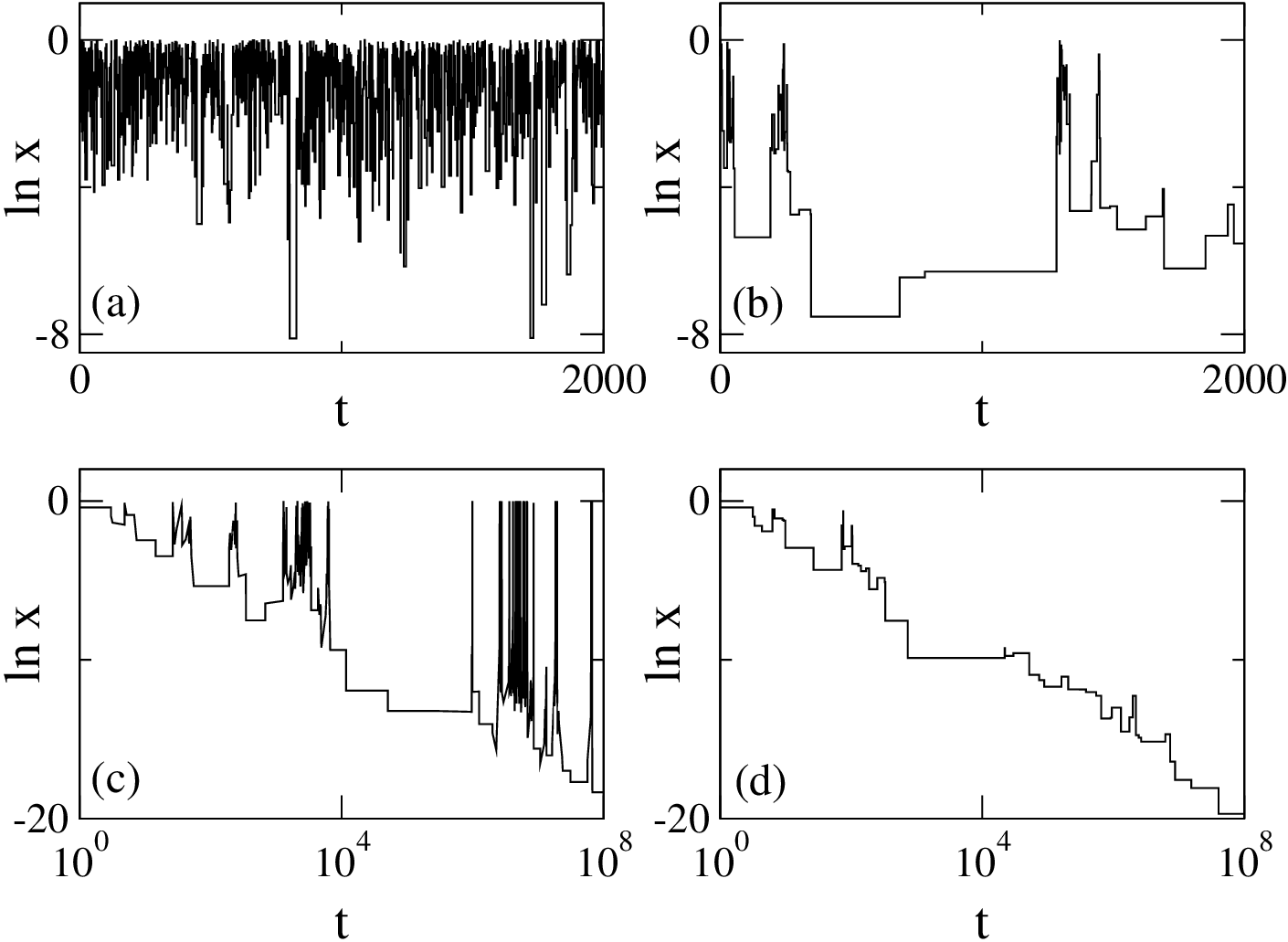}
\caption{Examples of trajectories for increasing values of $\gamma$.
(a) $\gamma=0.5$, (b) and (c) $\gamma=1.5$, (d) $\gamma=2.5$
($\eta=0$, $\nu_0=1$). (c) corresponds to the same data as (b), but plotted on a
logarithmic time scale.}
\label{fig-traj}
\end{figure}

{\it On-off intermittency: qualitative approach.}--
We show on Fig.~\ref{fig-traj} examples of trajectories corresponding
to different values of $\gamma$, using the function $\psi_{\mathrm{ex}}(u)$
defined in Eq.~(\ref{def-psi-ex}), with $\eta=0$,
and $\rho(x)=\gamma_0 x^{\gamma_0-1}$ ($0<x<1$).
We see on Fig.~\ref{fig-traj}(a) that for small values of $\gamma$,
the process remains in a fluctuating (or active) phase.
In contrast, for large enough values of $\gamma$, the dynamics
quickly converges to $x=0$, as seen on Fig.~\ref{fig-traj}(d).
For intermediate values of $\gamma$, on-off intermittency appears,
with a succession of active periods separated by quiet periods of time in which
$x$ remains very small [Fig.~\ref{fig-traj}(b)].
On a logarithmic time scale [Fig.~\ref{fig-traj}(c)],
one observes that lower and lower values of $x$ are progressively reached,
while bursts of activity where $x \approx 1$ remain present even at large time.

The fact that the process remains trapped
for long periods of time at small values of $x$ can be understood from
the transition rates (\ref{def-rates}). The average sojourn time $\tau_{\mathrm{s}}(x)$
at a given value $x$ can be evaluated by integrating over all possible escape paths:
\be
\frac{1}{\tau_{\mathrm{s}}(x)} = \int_0^{\infty} dx' \, W(x'|x).
\ee
From the small $x$ expansion of $\rho(x)$ given in Eq.~(\ref{rho-asympt}),
one finds $\tau_s(x) \sim \tau_1 \, x^{-\gamma_0}$ for $x \to 0$,
with $\tau_1 = [ \nu_0 c \int_0^{\infty} dv\, v^{\gamma_0-1} \psi(v) ]^{-1}$,
leading to a divergence of $\tau_s(x)$ in this limit.

As a first characterization of intermittency,
we determine the stationary distribution $p_\mathrm{st}(x)$.
Using the symmetry (\ref{def-psi}), one finds that a detailed balance
relation \cite{vanKampen}
$W(x'|x)\, p_\mathrm{st}(x) = W(x|x')\, p_\mathrm{st}(x')$
holds with respect to the distribution
\be \label{def-pst}
p_\mathrm{st}(x) = \frac{1}{Z}\, \rho(x)\, x^{-\gamma},
\ee
implying that $p_\mathrm{st}(x)$ is the stationary distribution.
We however need to check that the normalization constant
$Z = \int_0^{\infty} dx\, \rho(x)\, x^{-\gamma}$ is finite,
otherwise the distribution $p_\mathrm{st}(x)$ does not exist
(it cannot be normalized).
The convergence at the upper bound is ensured,
but the integral may not converge at its lower bound.
Using the small $x$ asymptotic behavior of $\rho(x)$, we find
$\rho(x)\, x^{-\gamma} \sim c\, x^{\gamma_0-\gamma-1}$,
so that the integral defining $Z$ converges only if $\gamma_0-\gamma >0$.
Hence the distribution $p_{\mathrm{st}}(x)$ is well-defined for $\gamma<\gamma_0$
but it becomes non-normalizable for $\gamma \ge \gamma_0$,
which indicates that the steady-state distribution should in this case be
a Dirac delta function at $x=0$ \cite{Aumaitre}.
For $\gamma \ge \gamma_0$, the time spent around $x=0$ thus completely
dominates the dynamics, which becomes nonstationary since the time
needed to reach the asymptotic delta distribution is infinite.

To sum up, the fact that $p_{\mathrm{st}}(x)$ becomes non-normalizable
accounts for the change of behavior between Fig.~\ref{fig-traj}(a) and (b)-(c).
We now need to understand the mechanism responsible for the different
behaviors observed in Fig.~\ref{fig-traj}(b)-(c) and (d).
In order to identify the value of $\gamma$ separating these two regimes,
we determine the average trend of the process on a logarithmic scale,
namely $\langle \Delta \ln x \rangle_x$ with $\Delta \ln x \equiv \ln x' -\ln x$.
The average is computed over $x'$ for a fixed value of $x$:
\be
\langle \Delta \ln x \rangle_x = \int_0^{\infty} dx' \, \phi(x'|x)\,
(\ln x' -\ln x)
\ee
where $\phi(x'|x)$ is the probability to jump from $x$ to $x'$, obtained by normalizing
the transition rate $W(x'|x)$:
\be \label{eq-phi}
\phi(x'|x) = \frac{W(x'|x)}{\int_0^{\infty} dx'' W(x''|x)}.
\ee
From the small $x$ behavior (\ref{rho-asympt}) of $\rho(x)$, we get
for small values of $x$ and $x'$
\be
\phi(x'|x) = \frac{1}{x} \, \tilde{\phi}\left(\frac{x'}{x}\right)
\ee
with the scaling function $\tilde{\phi}(u)= \nu_0 c \tau_1 \, u^{\gamma_0-1} \psi(u)$.
Using the symmetry of $\psi(u)$ given in Eq.~(\ref{def-psi}), one finds
\be \label{eq-dlnx}
\langle \Delta \ln x \rangle_x = \nu_0 c \tau_1
\int_1^{\infty} \frac{dv}{v} \ln v \,\psi(v) \left( v^{\gamma_0} - v^{\gamma-\gamma_0} \right).
\ee

\begin{figure}[t]
\centering\includegraphics[width=7.5cm,clip]{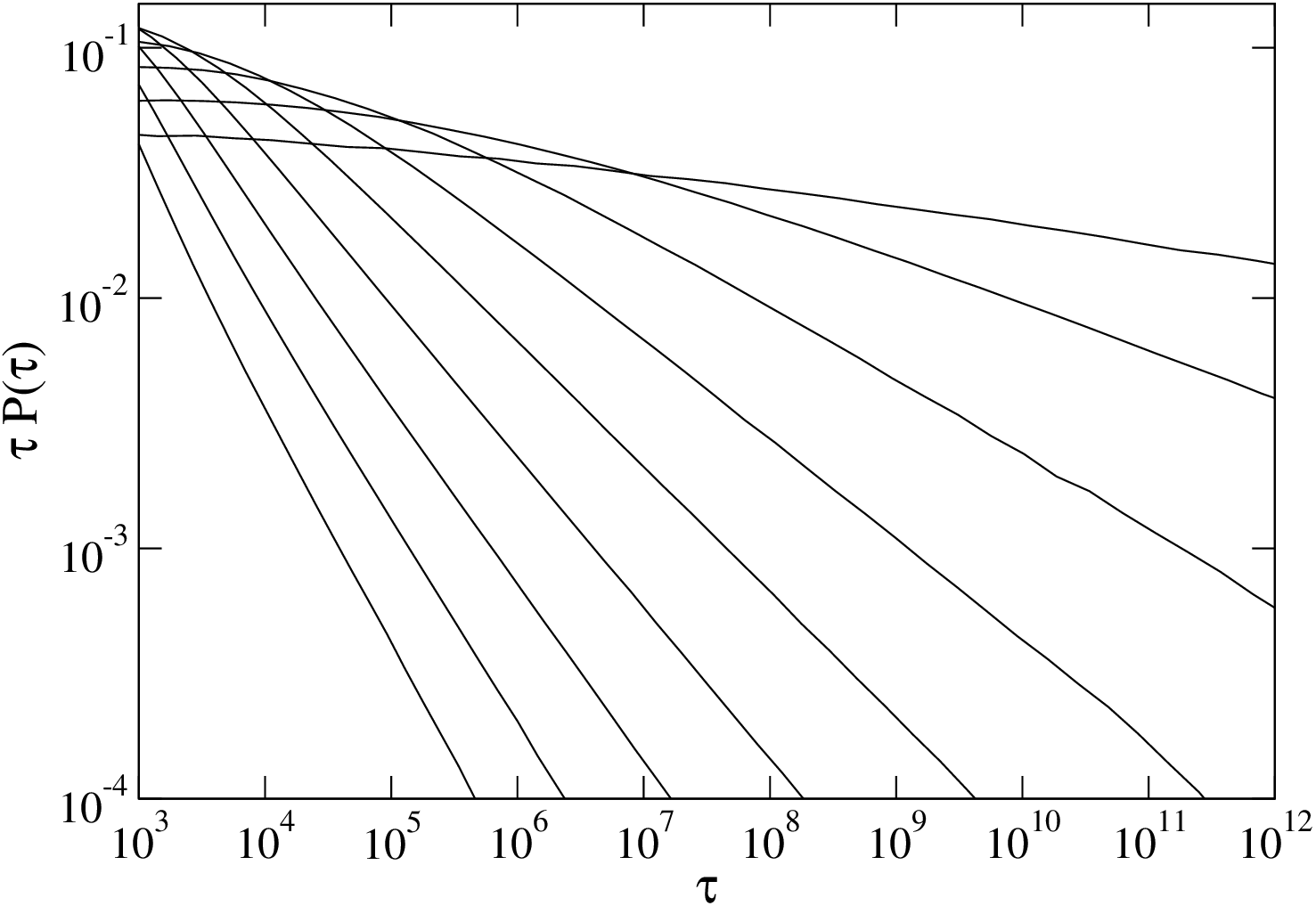}
\caption{Distribution $P(\tau)$ of the durations $\tau$ of the off-periods
for different values of $\gamma$. The distribution has been multiplied by
$\tau$ to enhance readability. From bottom to top:
$\gamma/\gamma_0=1.1$ to $1.9$ by steps of $0.1$ ($\eta=0$, $\nu_0=1$).}
\label{fig-dist}
\end{figure}

\noindent
We assume that the integral in Eq.~(\ref{eq-dlnx}) converges,
which is true in the example of $\psi_{\mathrm{ex}}(u)$ as soon as
$\gamma>\gamma_0-\eta$.
We first note that, in the considered small $x$ limit, the expression of
$\langle \Delta \ln x \rangle_x$ becomes independent of $x$.
Second, one readily sees from Eq.~(\ref{eq-dlnx}) that $\langle \Delta \ln x \rangle_x$
is positive for $\gamma_0 < \gamma < 2\gamma_0$, and negative for $\gamma > 2\gamma_0$.
Hence, when $\gamma > 2\gamma_0$, every step tends, on average, to reduce the value of $x$,
leading to a roughly monotonous convergence of $x$ towards $0$ [Fig.~\ref{fig-traj}(d)].
In contrast, for $\gamma_0 < \gamma < 2\gamma_0$, every step increases, on average,
the value of $x$, which favors finite values of $x$ instead of small ones.
Yet, rare negative steps play an important role due to the strong slowing down
of the dynamics for small values of $x$ [Fig.~\ref{fig-traj}(c)].
The competition between positive drift
and trapping at low values leads to the observed on-off intermittency phenomenon,
which is thus expected to exist over a finite range of the control parameter $\gamma$, namely
$\gamma_0 < \gamma < 2\gamma_0$.
Interestingly, the above analysis provides an interpretation of the
upper and lower bounds, $\gamma_{\mathrm{min}}=\gamma_0$ and $\gamma_{\mathrm{max}}=2\gamma_0$,
of the range over which on-off intermittency is present:
above $\gamma_{\mathrm{min}}$, kinetic trapping effects are strong close to $x=0$
(the distribution $p_{\mathrm{st}}(x)$ becomes non-normalizable),
while below $\gamma_{\mathrm{max}}$, the evolution is biased toward large values of $x$.
The existence of an overlap between these two ranges results in the presence
of on-off intermittency.
Note that $\gamma_{\mathrm{max}}$ plays a role similar to that of the instability
threshold in the standard on-off intermittency scenario.

\begin{figure}[t]
\centering\includegraphics[width=7cm,clip]{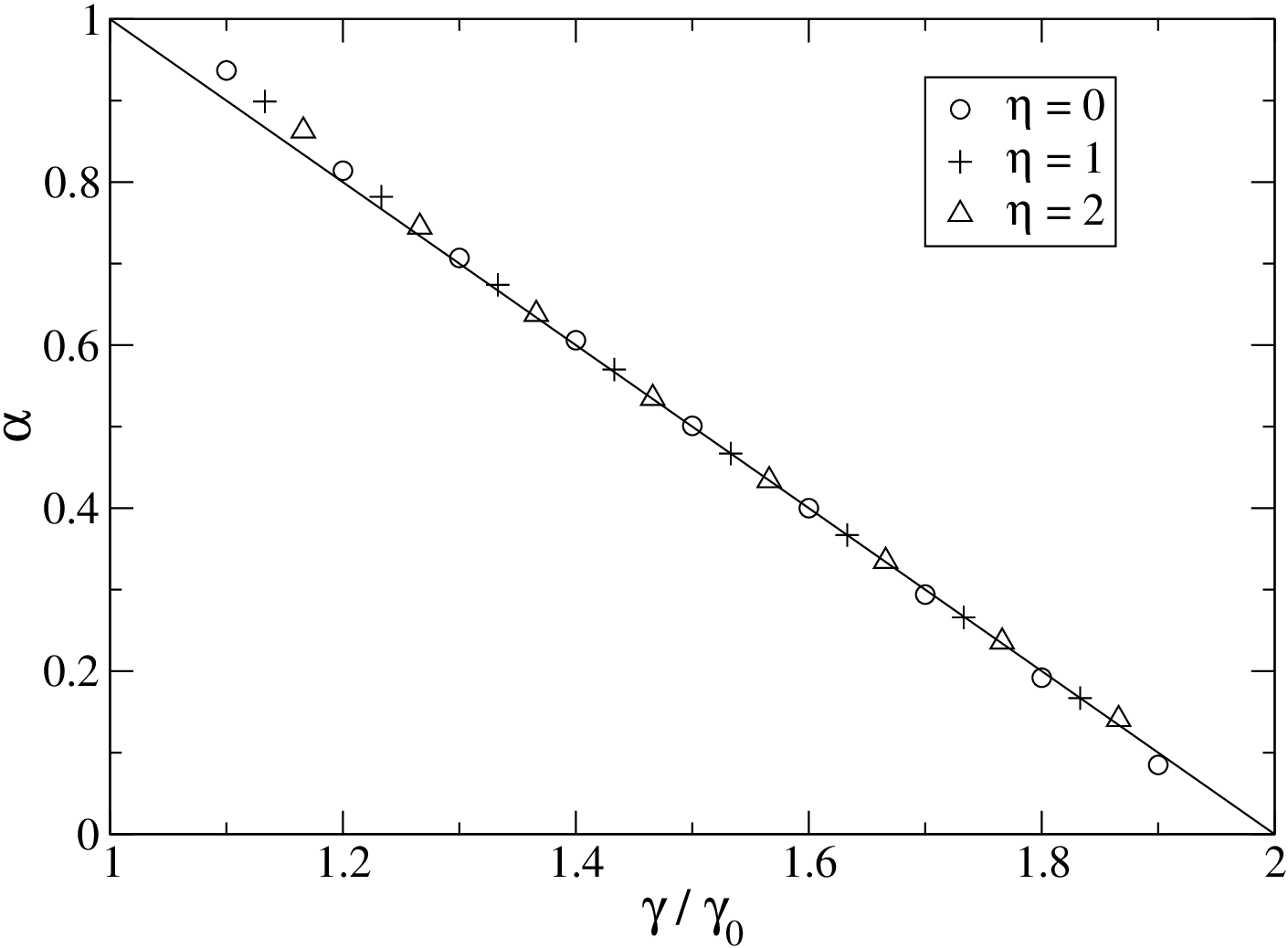}
\caption{Exponent $\alpha$ of the tail of the distribution of the
off-period durations as a function of the ratio $\gamma/\gamma_0$,
for different values of $\eta$.
The straight line is the prediction $\alpha=2-\gamma/\gamma_0$.}
\label{fig-alpha}
\end{figure}

{\it Distribution of off-periods.}--
In order to characterize more quantitatively the intermittent regime,
we have determined numerically the distribution $P(\tau)$ of the durations $\tau$
of the off-periods, by measuring the first return time to a given threshold $x_{\mathrm{th}}$
(we have checked that the results do not significantly depend on the precise
threshold value as long as $x_{\mathrm{th}} \approx 1$).
Simulations were performed using $\rho(x)=\gamma_0 x^{\gamma_0-1}$ and
the function $\psi_{\mathrm{ex}}(u)$
defined in Eq.~(\ref{def-psi-ex}), for different values of $\eta$.
The distribution $P(\tau)$ is plotted on Fig.~\ref{fig-dist}
for different values of $\gamma$ in the range $\gamma_0 < \gamma < 2\gamma_0$.
A power-law behavior is observed in the tail of the distribution, 
namely
\be \label{def-alpha}
P(\tau) \sim \frac{a}{\tau^{1+\alpha}}
\ee
for large $\tau$.
The exponent $\alpha$ is seen to be independent of $\eta$,
and varies almost linearly with the control
parameter $\gamma$ (see Fig.~\ref{fig-alpha}).
Small deviations from the power law can be seen for $\gamma$
close to $2\gamma_0$,
but the observed variation of the exponent $\alpha$ when varying the measurement
window over a reasonable range remains small,
at most of the order of the symbol size on Fig.~\ref{fig-alpha}.

We now give an analytical prediction of the exponent $\alpha$.
As the sojourn time $\tau_{\mathrm{s}}(x)$ at small values of $x$ becomes
very large, it is natural to assume (at variance with usual first return problems
of random walks)
that the return time $\tau$ is dominated by the largest sojourn time in the trajectory,
that is, the sojourn time $\tau_{\mathrm{s}}(x_{\mathrm{m}})$
at the smallest value of $x_{\mathrm{m}}$ reached.
Using the small $x$ expansion of $\tau_{\mathrm{s}}(x)$,
we thus have the scaling relation $\tau \sim \tau_1 \, x_{\mathrm{m}}^{-\gamma_0}$.
Here again, it is convenient to work with logarithmic variables.
We thus define the variable $z=-\gamma_0 \ln x$, from which the relation
$\tau \sim \tau_1 \, e^{z_{\mathrm{m}}}$ follows, where $z_{\mathrm{m}}$ is the maximal value
of $z$ along a trajectory between two crossings of the threshold
$z_{\mathrm{th}}=-\gamma_0 \ln x_{\mathrm{th}}$.
The distribution $P(\tau)$ can be deduced from the distribution
$\tilde{P}(z_{\mathrm{m}})$ through
$P(\tau) = \tilde{P}(z_{\mathrm{m}})\, dz_{\mathrm{m}}/d\tau$.
From the scaling relation $\tau \sim \tau_1 \, e^{z_{\mathrm{m}}}$, one obtains
\be \label{eq-Ptau}
P(\tau) \approx \frac{1}{\tau} \tilde{P}\left(\ln\frac{\tau}{\tau_1} \right).
\ee
In terms of number of steps (instead of real time), the trajectory of the logarithmic
variable $z$ is a random walk with a step statistics $\Phi(y)$
obtained from $\phi(x'|x)$ in the small $x$ limit [see Eq.~(\ref{eq-phi})]
through the correspondence $y=-\gamma_0 \ln(x'/x)$:
\be \label{def-Phi-rw}
\Phi(y) = \left\{\begin{array}{c}
b\, e^{-\mu y}, \qquad y>0 \\
b\, e^{\beta y}, \qquad \; \; y<0
\end{array}\right. 
\ee
where $\mu = 1+\eta/\gamma_0$, $\beta = (\gamma+\eta)/\gamma_0-1$
and $b=\beta\mu/(\beta+\mu)$.
We assume that $\mu>\beta$, or equivalently $\gamma<2\gamma_0$.
The distribution $\tilde{P}(z_{\mathrm{m}})$ can be determined from the auxiliary problem
of the random walk (\ref{def-Phi-rw}) with absorbing boundaries at $z=0$ and
$z=z_{\mathrm{m}}$.
We introduce the probability $Q(z_{\mathrm{th}},z_{\mathrm{m}})$ that the random walk,
starting from a position $z_{\mathrm{th}}$ ($0<z_{\mathrm{th}}<z_{\mathrm{m}}$),
is absorbed at the boundary $z_{\mathrm{m}}$.
The distribution $\tilde{P}(z_{\mathrm{m}})$ can be determined
by noting that the walks having a maximal value between $z_{\mathrm{m}}$
and $z_{\mathrm{m}}+dz_{\mathrm{m}}$
are precisely the walks that are absorbed by a boundary at $z_{\mathrm{m}}$,
but not by a boundary at $z_{\mathrm{m}}+dz_{\mathrm{m}}$, in the auxiliary problem.
It follows that
\be \label{eq-dQdz}
\tilde{P}(z_{\mathrm{m}}) = -\frac{\partial Q}{\partial z_{\mathrm{m}}}(z_{\mathrm{th}}, z_{\mathrm{m}}).
\ee
The probability $Q(z,z_{\mathrm{m}})$ satisfies the integral equation
\be \label{eq-Q}
Q(z,z_{\mathrm{m}}) = \int_{-z}^{z_{\mathrm{m}}-z} dy\, \Phi(y) \, Q(z+y,z_{\mathrm{m}})
+ \int_{z_{\mathrm{m}}-z}^{\infty} dy\, \Phi(y).
\ee
Looking for the solution of Eq.~(\ref{eq-Q}) as a sum
of exponentials of $z$, one finds
\be \label{eq-Q2}
Q(z,z_{\mathrm{m}}) = \frac{\beta}{\mu^2 e^{(\mu-\beta) z_{\mathrm{m}}} -\beta^2}
\left(\mu\, e^{(\mu-\beta) z} - \beta \right).
\ee
Using Eq.~(\ref{eq-dQdz}), we get for large $z_{\mathrm{m}}$
\be
\tilde{P}(z_{\mathrm{m}}) \approx K(z_{\mathrm{th}})\, e^{-(\mu-\beta) z_{\mathrm{m}}}
\ee
with $K(z_{\mathrm{th}}) = \beta(\mu-\beta) [\mu\, e^{(\mu-\beta) z_{\mathrm{th}}} - \beta]/\mu^2$.
From Eq.~(\ref{eq-Ptau}), we obtain for $P(\tau)$ the expression:
\be
P(\tau) \sim \frac{K(z_{\mathrm{th}})\, \tau_1^{\mu-\beta} }{\tau^{1+\mu-\beta}}, \qquad \tau \to \infty,
\ee
yielding $\alpha=\mu-\beta$, or in terms of $\gamma$,
\be \label{predict-alpha}
\alpha = 2-\frac{\gamma}{\gamma_0}.
\ee
This value, which is independent of $\eta$, is
in good agreement with the numerical results shown on Fig.~\ref{fig-alpha}.

The range $\gamma_0 < \gamma < 2\gamma_0$ over which intermittency appears
corresponds to $0<\alpha<1$, that is to a distribution $P(\tau)$
with an infinite mean value $\langle \tau \rangle$, as in standard
on-off intermittency for which $\alpha=1/2$.
If $\gamma < \gamma_0$, the average duration of off-periods is finite, while
if $\gamma > 2\gamma_0$, the process becomes completely trapped in the off state,
without any return to the threshold value, so that the distribution $P(\tau)$
can no longer be defined.
We also note that the exponent $\alpha$ is independent of the threshold value
$z_{\mathrm{th}}$ [which only appears in the prefactor $K(z_{\mathrm{th}})$],
as also observed numerically.

{\it Discussion.}-- In this note, we have considered a simple model
exhibiting a finite range of control parameter
over which on-off intermittency is present,
with a continuously varying exponent $\alpha$ characterizing the distribution
of the duration of off-periods. The mechanism at the origin of
the intermittency phenomenon and of the power-law
distribution $P(\tau)$ differs from that found in standard on-off intermittency.
In the standard case,
the statistics of off-periods is related to the distribution of return times
of unbiased homogeneous random walks (whence the exponent $\alpha=1/2$ follows),
while in the present model,
on-off intermittency results from the competition of bias and slowing down
effects, leading to a non-trivial statistics with a tunable exponent
$\alpha$, and to an extended range of intermittency.

Further work is needed to identify more clearly the relation between
the present model and the standard scenario of on-off intermittency.
For instance, it would be of interest to make contact with dynamical systems
theory, for instance by finding models having a behavior similar
to the present one, but being defined by a stochastic differential equation
or by a chaotic map.
Another open question is the physical interpretation of the
control parameter $\gamma$, which is a phenomenological parameter
at this stage. As a first attempt in this direction, we note
that the present model can be mapped, for $\eta=0$, onto
the Barrat-M\'ezard model \cite{BM95,Bertin03}, a simple model exhibiting
aging dynamics. In this mapping, $\gamma$ is found to be proportional
to the inverse temperature and thus acquires a simple physical interpretation.
It would be interesting to find other examples of physical realizations of
the present model.


\begin{thebibliography}{99}


\bibitem{Hammer94}
P.~W. Hammer, N. Platt, S.~M. Hammel, J.~F. Heagy, and B. D. Lee,
Phys. Rev. Lett. \textbf{73}, 1095 (1994).

\bibitem{Rodelsperger95}
F. R\"odelsperger, A. Cenys, and H. Benner, Phys. Rev. Lett. \textbf{75},
2594 (1995).

\bibitem{Feng98}
D.~L. Feng, C.~X. Yu, J.~L. Xie, and W.~X. Ding, Phys. Rev. E \textbf{58},
3678 (1998).

\bibitem{John99}
T. John, R. Stannarius, and U. Behn, Phys. Rev. Lett. \textbf{83},
749 (1999).

\bibitem{Vella03}
A. Vella, A. Setaro, B. Piccirillo, and E. Santamato, Phys. Rev. E
\textbf{67}, 051704 (2003).

\bibitem{Huerta08}
G. Huerta-Cuellar, A.~N. Pisarchik, and Y.~O. Barmenkov,
Phys. Rev. E {\bf 78}, 035202 (2008).

\bibitem{Brokmann03}
X. Brokmann \textsl{et al.}, Phys. Rev. Lett. \textbf{90}, 120601 (2003).

\bibitem{Zumofen04}
G. Zumofen, J. Hohlbein and C.~G. Hubner, Phys. Rev. Lett. \textbf{93},
260601 (2004).
 
\bibitem{Fujisaka1985}
H. Fujisaka and T. Yamada, Prog. Theor. Phys. \textbf{74}, 918 (1985).

\bibitem{Platt1993}
N. Platt, E.~A. Spiegel, and C. Tresser, Phys. Rev. Lett. \textbf{70},
279 (1993).

\bibitem{Heagy1994}
J.~F. Heagy, N. Platt, and S.~M. Hammel, Phys. Rev. E \textbf{49},
1140 (1994).

\bibitem{Aumaitre}
S. Auma\^{\i}tre, F. P\'etr\'elis and K. Mallick, Phys. Rev. Lett. {\bf 95},
064101 (2005);
S. Auma\^{\i}tre, K. Mallick, and F. P\'etr\'elis, J. Stat. Phys. {\bf 123}, 909 (2006).

\bibitem{Feller}
W. Feller, \emph{An introduction to probability theory and its applications},
Vol.~I, $3^{\mathrm{rd}}$ ed. (Wiley, New-York, 1968).

\bibitem{Kuno01}
M. Kuno, D.~P. Fromm, H.~F. Hamann, A. Gallagher, and D.~J. Nesbitt,
J. Chem. Phys. {\bf 115}, 1028 (2001).

\bibitem{Kuno03}
M. Kuno, D.~P. Fromm, S.~T. Johnson, A. Gallagher, and D.~J. Nesbitt,
Phys. Rev. B {\bf 67}, 125304 (2003).

\bibitem{Divoux}
T. Divoux, E. Bertin, V. Vidal, J.-C. G\'eminard, Phys. Rev. E {\bf 79}, 056204 (2009).

\bibitem{fractal-noise95}
M. Ding and W. Yang, Phys. Rev. E {\bf 52}, 207 (1995).

\bibitem{fractal-noise96}
H.~L. Yang, Z.~Q. Huang, and E.~J. Ding, Phys. Rev. E {\bf 54}, 3531 (1996).

\bibitem{vanKampen}
N.~G. Van Kampen, ``Stochastic Processes in Physics and Chemistry''
(North Holland, 1992).

\bibitem{BM95}
A. Barrat and M. M\'ezard, J. Phys. I (France) {\bf 5}, 941 (1995).

\bibitem{Bertin03}
E. Bertin, J. Phys. A: Math. Gen. {\bf 36}, 10683 (2003).



\end{thebibliography}
\end{document}